# BENCHMARKING MACHINE LEARNING TECHNIQUES FOR SOFTWARE DEFECT DETECTION

Saiqa Aleem[1], Luiz Fernando Capretz[1] and Faheem Ahmed[2]

[1] Western University, Department of Electrical & Computer Engineering,
London, Ontario, Canada, N6A 5B9
[2] Thompson Rivers University, Department of Computing Science,
Kamloops, British Columbia, Canada, V2C 6N6

## ABSTRACT

*Machine Learning approaches are good in solving problems that have less information. In most cases, the software domain problems characterize as a process of learning that depend on the various circumstances and changes accordingly. A predictive model is constructed by using machine learning approaches and classified them into defective and non-defective modules. Machine learning techniques help developers to retrieve useful information after the classification and enable them to analyse data from different perspectives. Machine learning techniques are proven to be useful in terms of software bug prediction. This study used public available data sets of software modules and provides comparative performance analysis of different machine learning techniques for software bug prediction. Results showed most of the machine learning methods performed well on software bug datasets.*

## 1. INTRODUCTION

The advancement in software technology causes an increase in the number of software products, and their maintenance has become a challenging task. More than half of the life cycle cost for a software system includes maintenance activities. With the increase in complexity in software systems, the probability of having defective modules in the software systems is getting higher [1]. It is imperative to predict and fix the defects before it is delivered to customers because the software quality assurance is a time consuming task and sometimes does not allow for complete testing of the entire system due to budget issue. Therefore, identification of a defective software module can help us in allocating limited and resources effectively. A defect in a software system can also be named a bug.

A bug indicates the unexpected behaviour of system for some given requirements. The unexpected behaviour is identified during software testing and marked as a bug. A software bug can be referred to as" Imperfection in software development process that would cause software to fail to meet the desired expectation" [2]. Moreover, the finding of defects and correcting those results in expensive software development activities [3]. It has been observed that a small number of modules contain the majority of the software bugs [4, 5]. Thus, timely identification of software bugs facilitates the testing resources allocation in an efficient manner and enables developers to improve the architectural design of a system by identifying the high risk segments of the system [6, 7, 8].

 11

Machine learning techniques can be used to analyse data from different perspectives and enable developers to retrieve useful information. The machine learning techniques that can be used to detect bugs in software datasets can be classification and clustering. Classification is a data mining and machine learning approach, useful in software bug prediction. It involves categorization of software modules into defective or non-defective that is denoted by a set of software complexity metrics by utilizing a classification model that is derived from earlier development projects data [9]. The metrics for software complexity may consist of code size [10], McCabe's cyclomatic complexity [11] and Halstead's Complexity [12].

Clustering is a kind of non-hierarchal method that moves data points among a set of clusters until similar item clusters are formed or a desired set is acquired. Clustering methods make assumptions about the data set. If that assumption holds, then it results into a good cluster. But it is a trivial task to satisfy all assumptions. The combination of different clustering methods and by varying input parameters may be beneficial. Association rule mining is used for discovering frequent patterns of different attributes in a dataset. The associative classification most of the times provides a higher classification as compared to other classification methods.

This paper explores the different machine learning techniques for software bug detection and provides a comparative performance analysis between them. The rest of the paper is organized as follows: Section II provides a related work on the selected research topic; Section III discusses the different selected machine learning techniques, data pre-process and prediction accuracy indicators, experiment procedure and results; Section VI provides the discussion about comparative analysis of different methods; and Section V concludes the research.

## 2. RELATED WORK

Lessmann et al. [13] proposed a novel framework for software defect prediction by benchmarking classification algorithms on different datasets and observed that their selected classification methods provide good prediction accuracy and supports the metrics based classification. The receiver operating characteristics curve (AUC) is used for comparison. Actually, AUC represents the objective indicator of predictive accuracy and it is most informative within a benchmarking context [14, 15]. Especially for comparative study in software bug detection, it is recommended to use AUC as primary accuracy indicator because it separates predictive performance from cost distributions and class, and they are actually project specific characteristics that may be subject to change and unknown. Therefore, there is a potential for AUC-based evaluation to significantly improve convergence across studies. In particular, of RndFor for defect prediction previous findings regarding the efficacy [16] were confirmed. The results of the experiments showed that there is no significant difference in the performance of different classification algorithms. The study covered only classification model for software bug prediction.

Sharma and Jain [17] explored the WEKA approach for decision tree classification algorithms. They characterized specific approach for classification and developed method for WEKA in order to utilize the implementation of different datasets. The high rate of accuracy is presented and achieved by each decision tree. It correctly classify data into its related instances. The proposed approach can be used in banking, medical and various areas. Their proposed method is generic one not especially for software bug prediction. Various machine learning approaches such as Artificial Neural Network (ANN), Bayesian Belief Network (BBN), Decision Tree, clustering and SVM are some techniques which are generally used for fault prediction in software. Elish and Elish [18] proposed a software prediction model by utilizing SVM approach. A comparative analysis was also performed for SVM against four NASA datasets including eight machine learning models. Guo et al., [19] also used NASA software bugs datasets and utilized ensemble approach (Random Forest) to predict non-defective software components and also compared its performance against other existing machine leaning approaches. Ghouti *et al.*, [20] proposed a model based on Probabilistic Neural Network (PNN) and SVM for fault prediction and used PROMISE datasets for evaluation. This research work suggested that predictive performance of PNN is better than SVM for any size of datasets.

Khoshgoftaar *et al.* [21] also used one of the machine learning approach i.e. Neural Network to find out that either software module is defective or not and performed experiment on large tele-communication. They did a comparative analysis between NN and other approaches and concluded that NN performed well in bug prediction as compared to other approaches. Kaur and Pallavi [22] also discussed the utilization of numerous machine learning approaches for example classification, clustering, regression, association and regression in software defect prediction but did not provide the comparative performance analysis of techniques. Okutan and Yildiz [23] and Fenton *et al.* [24] and also predict bugs in software modules by using Bayesian Network approach. Okutan and Yildiz. [23] used PROMISE data repository and concluded that most effective metrics for software are response for class, lines of code and lack of coding quality. Wang *et al.* [25] provided a comparative study of only ensemble classifiers for software bug prediction.

Most of the existed studies on software defect prediction are limited in performing comparative analysis of all the methods of machine learning. Some of them used few methods and provides the comparison between them and others just discussed or proposed a method based on existing machine learning techniques by extending them [26, 27].

## 3. MACHINE LEARNING TECHNIQUES FOR SOFTWARE BUG DETECTION

In this paper, a comparative performance analysis of different machine learning techniques is explored for software bug prediction on public available data sets. Machine learning techniques are proven to be useful in terms of software bug prediction. The data from software repository contains lots of information in assessing software quality; and machine learning techniques can be applied on them in order to extract software bugs information. The machine learning techniques are classified into two broad categories in order to compare their performance; such as supervised learning versus unsupervised learning. In supervised learning algorithms such as ensemble classifier like bagging and boosting, Multilayer perceptron, Naive Bayes classifier, Support vector machine, Random Forest and Decision Trees are compared. In case of unsupervised learning methods like Radial base network function, clustering techniques such as K-means algorithm, K nearest neighbour are compared against each other.

The brief description of each one is as follow:

### 3.1.1 Decision Tree

Decision trees classify software defective modules by using a series of rule [28]. The decision tree has basic components such as the decision node, branches and leaves. Input space within decision tree is divided into mutually exclusive regions and a value or an action or a label is assigned to each region to characterize its data points. The mechanism of decision tree is transparent and decision tree structure can be follow to see how the decision is made. Most of the decision trees construction algorithm consists of two phases. In the first phase, vary large size tree is constructed and then the tree is pruned in the second step to avoid over fitting issue. Then the pruned tree is utilized for classification purpose.

### 3.1.2 Ensemble Classifier (Bagging and Boosting)

Ensemble Classifier integrates multiple classifier to build a model for classification and helps in improving the defect prediction performance. The main idea is to improve the overall performance of prediction by combining set of learning models. The Bagging [29] (Bootstrap AGGregatING) is one of the ensemble classifier and mainly constructs each ensemble member by using different datasets. Then the predictions are made by combining their average or votes over a label of class. Bagging build a combined model results in better performance than one single model. Another ensemble method is Boosting and Adaboost [30] is one of the well-known algorithm of Boosting family. It usually train new model in each round and multiple iterations with different example weights are performed. The increment in the weight of incorrectly classified classes will be done, so this over fitting counts more heavily in the next iteration. In this way, the series of classifier complement each other and they are combined together by voting.

### 3.1.3 Random Forest

Random Forest [31] is also another approach under ensemble classifier. In the construction of decision tree a random choice of attributes is involved. A simple algorithm is used in the construction of individual tree. Pruning process is not performed at each node of decision tree and sampling of attributes is randomly performed. The unlabelled example classified based on majority of voting [32]. Random forest has one important advantage that it is fast and is able to handle large number of input attributes.

### 3.1.4 Naïve Bayes Classifiers (NB)

The Naïve Bayes classifier [33] is based on Bayes rule of conditional probability. It analysis each attribute individually and assumes that all of them are independent and important.

### 3.1.5 Support Vector Machine (SVM)

A support vector machine (or SVM) [34, 35] utilizes non-linear mapping for original training data to transform it into higher dimension. Then it searches for optimal linear hyper plane for separation. The hyper-plane can be found using margins and support vectors. SVM is used for classification purpose and based on supervised learning.

### 3.1.6 Multi-layer Perceptron (MLP)

A multilayer perceptron (MLP) [36] is a supervised learning approach and comprised of feedforward artificial neural network model. The sets of input data in this approach map onto a set of appropriate outputs. A MLP comprised of directed graph of multiple layers of nodes, and they are fully connected to the next one within each node. Each input node is called as neuron with a nonlinear activation function. The sigmoidal units of hidden layer learn to approximate the functions. For training purpose, MLP utilizes a technique called backpropagation.

### 3.1.7 Radial Basis Function Networks

Radial basis function (RBF) [37] Networks uses the approximation theory of function. It is different from MLP because it has feed forward networks of two layers. The radial basis functions are implemented within hidden nodes and output nodes utilizes linear summation functions. The learning and training is very fast in RBF networks.

### 3.1.8 Clustering

Clustering is classified under unsupervised learning approach because no class labels are provided. The data is grouped together on the basis of their similarity. Groups with similar data points are put together in clusters. It is a process defining set of meaningful sub-classes called clusters based on their similarities. K-mean [38] clustering is based on non-hierarchical clustering procedure and item are moved within sets of clusters until the desired set is reached. K- Nearest neighbors is also another example of clustering under unsupervised learning.

### 3.2 Datasets & Pre-Processing

The datasets from PROMISE data repository [39] were used in the experiments. Table 1 shows the information about datasets. The datasets were collected from real software projects by NASA and have many software modules. We used public domain datasets in the experiments as this is a benchmarking procedure of defect prediction research, making easier for other researcher to compare their techniques [13, 8]. Datasets used different programming languages and code metrics such as Halstead's complexity, code size and McCabe's cyclomatic complexity etc. Experiments were performed by such a baseline.

Waikato Environment for Knowledge Analysis (WEKA) [40] tool was used for experiments. It is an open source software consisting of a collection of machine learning algorithms in java for different machine learning tasks. The algorithms are applied directly to different datasets. Pre-processing of datasets has been performed before using them in the experiments. Missing values were replaced by the attribute values such as means of attributes because datasets only contain numeric values. The attributes were also discretized by using filter of *Discretize* (10-bin discretization) in WEKA software. The data file normally used by WEKA is in ARFF file format, which consists of special tags to indicate different elements in the data file (foremost: attribute names, attribute types, and attribute values and the data).

### 3.3 Performance Indicators

For comparative study, performance indicators such as accuracy, mean absolute error and F-measure based on precision and recall were used. Accuracy can be defined as the total number of correctly identified bugs divided by the total number of bugs, and is calculated by the equations listed below:

Accuracy = (TP + TN) / (TP+TN+FP+FN)  (1)

Accuracy (%) = (correctly classified software bugs/ Total software bugs) * 100  (2)

Precision is a measure of correctness and it is a ratio between correctly classified software bugs and actual number of software bugs assigned to their category. It is calculated by the equation below:

Precision = TP / (TP+FP)  (3)

Table 1. Datasets information

|  | CM1 | JM1 | KC1 | KC2 | KC3 | MC1 | MC2 | MW1 | PC1 | PC2 | PC3 | PC4 | PC5 | AR1 | AR6 |
|---|---|---|---|---|---|---|---|---|---|---|---|---|---|---|---|
| Language | C | C | C++ | C++ | Java | C++ | C | C | C | C | C | C | C++ | C | C |
| LOC | 20k | 315k | 43k | 18k | 18k | 63k | 6k | 8k | 40k | 26k | 40k | 36k | 164k | 29k | 29 |
| Modules | 505 | 10878 | 2107 | 522 | 458 | 9466 | 161 | 403 | 1107 | 5589 | 1563 | 1458 | 17186 | 121 | 101 |
| Defects | 48 | 2102 | 325 | 105 | 43 | 68 | 52 | 31 | 76 | 23 | 160 | 178 | 516 | 9 | 15 |

Table 2. Performance of different machine learning methods with cross validation test mode based on accuracy

| Datasets | Supervised learning | | | | | | | | Unsupervised learning | | |
|---|---|---|---|---|---|---|---|---|---|---|---|
|  | Naye Bayes | MLP | SVM | Ada Boost | Bagging | Decision Trees | Random Forest | J48 | KNN | RBF | K-means |
| AR1 | 83.45 | 89.55 | 91.97 | 90.24 | 92.23 | 89.32 | 90.56 | 90.15 | 65.92 | 90.33 | 90.02 |
| AR6 | 84.25 | 84.53 | 86.00 | 82.70 | 85.18 | 82.88 | 85.39 | 83.21 | 75.13 | 85.38 | 83.65 |
| CM1 | 84.90 | 89.12 | 90.52 | 90.33 | 89.96 | 89.22 | 89.40 | 88.71 | 84.24 | 89.70 | 86.58 |
| JM1 | 81.43 | 89.97 | 81.73 | 81.70 | 82.17 | 81.78 | 82.09 | 80.19 | 66.89 | 81.61 | 77.37 |
| KC1 | 82.10 | 85.51 | 84.47 | 84.34 | 85.39 | 84.88 | 85.39 | 84.13 | 82.06 | 84.99 | 84.03 |
| KC2 | 84.78 | 83.64 | 82.30 | 81.46 | 83.06 | 82.65 | 82.56 | 81.29 | 79.03 | 83.63 | 80.99 |
| KC3 | 86.17 | 90.04 | 90.80 | 90.06 | 89.91 | 90.83 | 89.65 | 89.74 | 60.59 | 89.87 | 87.91 |
| MC1 | 94.57 | 99.40 | 99.26 | 99.27 | 99.42 | 99.27 | 99.48 | 99.37 | 68.58 | 99.27 | 99.48 |
| MC2 | 72.53 | 67.97 | 72.00 | 69.46 | 71.54 | 67.21 | 70.50 | 69.75 | 64.49 | 69.51 | 69.00 |
| MW1 | 83.63 | 91.09 | 92.19 | 91.27 | 92.06 | 90.97 | 91.29 | 91.42 | 81.77 | 91.99 | 87.90 |
| PC1 | 88.07 | 93.09 | 93.09 | 93.14 | 93.79 | 93.36 | 93.54 | 93.53 | 88.22 | 93.13 | 92.07 |
| PC2 | 96.96 | 99.52 | 99.59 | 99.58 | 99.58 | 99.58 | 99.55 | 99.57 | 75.25 | 99.58 | 99.21 |
| PC3 | 46.87 | 87.55 | 89.83 | 89.70 | 89.38 | 89.60 | 89.55 | 88.14 | 64.07 | 89.76 | 87.22 |
| PC4 | 85.51 | 89.11 | 88.45 | 88.86 | 89.53 | 88.53 | 89.69 | 88.36 | 56.88 | 87.27 | 86.72 |
| PC5 | 96.93 | 97.03 | 97.23 | 96.84 | 97.59 | 97.01 | 97.58 | 97.40 | 66.77 | 97.15 | 97.33 |
| Mean | 83.47 | 89.14 | 89.29 | 88.59 | 89.386 | 88.47 | 89.08 | 88.33 | 71.99 | 88.87 | 87.29 |

Recall is a ratio between correctly classified software bugs and software bugs belonging to their category. It represents the machine learning method's ability of searching extension and is calculated by the following equation.

Recall = TP / (TP + FN)  (4)

F-measure is a combined measure of recall and precision, and is calculated by using the following equation. The higher value of F-measure indicates the quality of machine learning method for correct prediction.

F = (2 * precision * recall) / (Precision + recall)  (5)

Table 3. Performance of different machine learning methods with cross validation test mode based on mean absolute error

| Datasets | Supervised learning | | | | | | | | Unsupervised learning | | |
|---|---|---|---|---|---|---|---|---|---|---|---|
| | NayeBayes | MLP | SVM | AdaBoost | Bagging | Decision Trees | Random Forest | J48 | KNN | RBF | K-means |
| AR1 | 0.17 | 0.11 | 0.08 | 0.12 | 0.13 | 0.12 | 0.13 | 0.13 | 0.32 | 0.13 | 0.11 |
| AR6 | 0.17 | 0.19 | 0.13 | 0.22 | 0.24 | 0.25 | 0.22 | 0.23 | 0.25 | 0.22 | 0.17 |
| CM1 | 0.16 | 0.16 | 0.10 | 0.16 | 0.16 | 0.20 | 0.16 | 0.17 | 0.16 | 0.17 | 0.14 |
| JM1 | 0.19 | 0.27 | 0.18 | 0.27 | 0.25 | 0.35 | 0.25 | 0.26 | 0.33 | 0.28 | 0.23 |
| KC1 | 0.18 | 0.21 | 0.15 | 0.22 | 0.20 | 0.29 | 0.19 | 0.20 | 0.18 | 0.23 | 0.17 |
| KC2 | 0.16 | 0.22 | 0.17 | 0.22 | 0.22 | 0.29 | 0.22 | 0.23 | 0.21 | 0.23 | 0.21 |
| KC3 | 0.15 | 0.12 | 0.09 | 0.14 | 0.14 | 0.17 | 0.14 | 0.13 | 0.39 | 0.15 | 0.12 |
| MC1 | 0.06 | 0.01 | 0.01 | 0.01 | 0.01 | 0.03 | 0.01 | 0.01 | 0.31 | 0.01 | 0.01 |
| MC2 | 0.27 | 0.32 | 0.28 | 0.39 | 0.37 | 0.40 | 0.35 | 0.32 | 0.35 | 0.41 | 0.31 |
| MW1 | 0.16 | 0.11 | 0.08 | 0.12 | 0.12 | 0.15 | 0.12 | 0.12 | 0.18 | 0.12 | 0.13 |
| PC1 | 0.11 | 0.11 | 0.07 | 0.11 | 0.10 | 0.14 | 0.09 | 0.10 | 0.12 | 0.12 | 0.08 |
| PC2 | 0.03 | 0.01 | 0.00 | 0.01 | 0.01 | 0.02 | 0.01 | 0.01 | 0.18 | 0.01 | 0.01 |
| PC3 | 0.51 | 0.14 | 0.10 | 0.16 | 0.15 | 0.21 | 0.15 | 0.15 | 0.36 | 0.18 | 0.13 |
| PC4 | 0.14 | 0.12 | 0.11 | 0.15 | 0.14 | 0.16 | 0.14 | 0.12 | 0.43 | 0.20 | 0.13 |
| PC5 | 0.04 | 0.03 | 0.03 | 0.04 | 0.03 | 0.06 | 0.03 | 0.03 | 0.33 | 0.05 | 0.03 |
| Mean | 0.16 | 0.14 | 0.10 | 0.15 | 0.15 | 0.18 | 0.14 | 0.14 | 0.27 | 0.16 | 0.13 |

## 3.4 Experiment Procedure & Results

For comparative performance analysis of different machine learning methods, we selected 15 software bug datasets and applied machine learning methods such as NaiveBayes, MLP, SVM, AdaBoost, Bagging, Decision Tree, Random Forest, J48, KNN, RBF and K-means. We employed WEKA tool for the implementation of experiments. The 10- fold cross validation test mode was selected for the experiments.

**Experiment Procedure:**

**Input:**

*i) The software bug repository datasets:*
 D= {AR1, AR6, CM1, JM1, KC1, KC2, KC3, MC1, MC2, MW1, PC1, PC2, PC3, PC4, PC5}
*ii) Selected machine learning methods*
M = {Nayes Bayes, MLP, SVM, AdaBoost, Bagging, Decision Tree, Random Forest, J48, KNN, RBF, K-means}

**Data pre-process:**
a)      Apply Replace missing values to D
b)      Apply Discretize to D
**Test Model - cross validation (10 folds):**
for each D do for each M do
        Perform cross-validation using 10-folds
end for
Select accuracy
Select Mean Absolute Error (MAE) Select F-measure end for
**Output:**
a)      Accuracy
b)      Mean Absolute Error
c)      F-measure

Table 4. Performance of different machine learning methods with cross validation test mode based on F-measure

| Datasets | Supervised learning | | | | | | | | Unsupervised learning | | |
|---|---|---|---|---|---|---|---|---|---|---|---|
| | NayeBayes | MLP | SVM | AdaBoost | Bagging | Decision Trees | Random Forest | J48 | KNN | RBF | K-means |
| AR1 | 0.90 | 0.94 | 0.96 | 0.95 | 0.96 | 0.94 | 0.96 | 0.95 | 0.79 | 0.95 | 0.94 |
| AR6 | 0.90 | 0.91 | 0.93 | 0.90 | 0.92 | 0.90 | 0.92 | 0.90 | 0.84 | 0.92 | 0.90 |
| CM1 | 0.91 | 0.94 | 0.95 | 0.95 | 0.95 | 0.94 | 0.94 | 0.94 | 0.91 | 0.95 | 0.93 |
| JM1 | 0.89 | 0.90 | 0.90 | 0.90 | 0.90 | 0.90 | 0.90 | 0.88 | 0.80 | 0.90 | 0.86 |
| KC1 | 0.90 | 0.92 | 0.92 | 0.91 | 0.92 | 0.92 | 0.92 | 0.91 | 0.89 | 0.92 | 0.91 |
| KC2 | 0.90 | 0.90 | 0.90 | 0.88 | 0.90 | 0.89 | 0.89 | 0.88 | 0.86 | 0.90 | 0.88 |
| KC3 | 0.91 | 0.94 | 0.95 | 0.95 | 0.95 | 0.95 | 0.94 | 0.94 | 0.72 | 0.95 | 0.93 |
| MC1 | 0.97 | 1.00 | 1.00 | 1.00 | 1.00 | 1.00 | 1.00 | 1.00 | 0.81 | 1.00 | 1.00 |
| MC2 | 0.82 | 0.78 | 0.82 | 0.80 | 0.81 | 0.77 | 0.80 | 0.78 | 0.76 | 0.81 | 0.77 |
| MW1 | 0.90 | 0.95 | 0.96 | 0.95 | 0.96 | 0.95 | 0.95 | 0.95 | 0.89 | 0.96 | 0.93 |
| PC1 | 0.94 | 0.97 | 0.96 | 0.96 | 0.97 | 0.97 | 0.97 | 0.97 | 0.94 | 0.96 | 0.96 |
| PC2 | 0.99 | 1.00 | 1.00 | 1.00 | 1.00 | 1.00 | 1.00 | 1.00 | 0.90 | 1.00 | 1.00 |
| PC3 | 0.60 | 0.94 | 0.95 | 0.95 | 0.94 | 0.95 | 0.94 | 0.94 | 0.77 | 0.95 | 0.93 |
| PC4 | 0.92 | 0.94 | 0.94 | 0.94 | 0.94 | 0.93 | 0.94 | 0.93 | 0.72 | 0.93 | 0.92 |
| PC5 | 0.98 | 0.99 | 0.99 | 0.98 | 0.99 | 0.98 | 0.99 | 0.99 | 0.80 | 0.99 | 0.99 |
| Mean | 0.89 | 0.93 | 0.942 | 0.93 | 0.94 | 0.93 | 0.93 | 0.93 | 0.82 | 0.93 | 0.92 |

## 3.5 Experiment Results

Table 2, 3 & 4 show the results of the experiment. Three parameters were selected in order to compare them such as Accuracy, Mean absolute error and F-measure. In order to compare the selected algorithms the mean was taken for all datasets and results are shown in Figures 1-3.

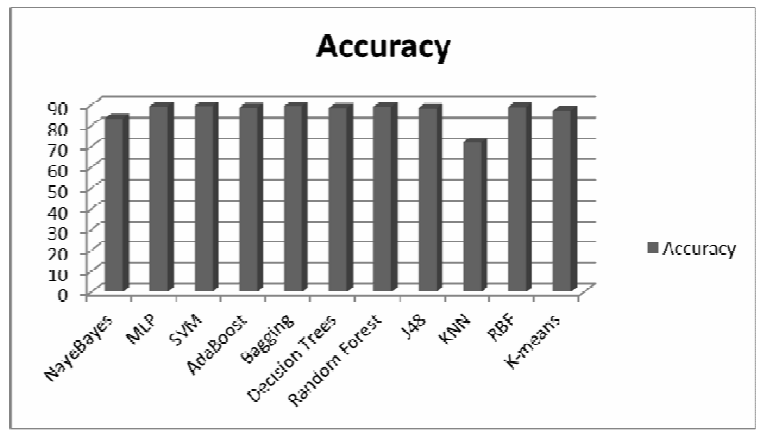

Figure 1. Accuracy results for selected machine learning methods

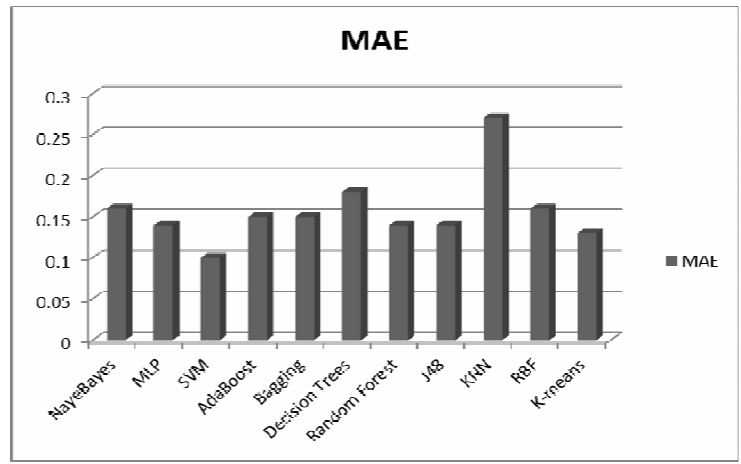

Figure 2. MAE results for selected machine learning methods

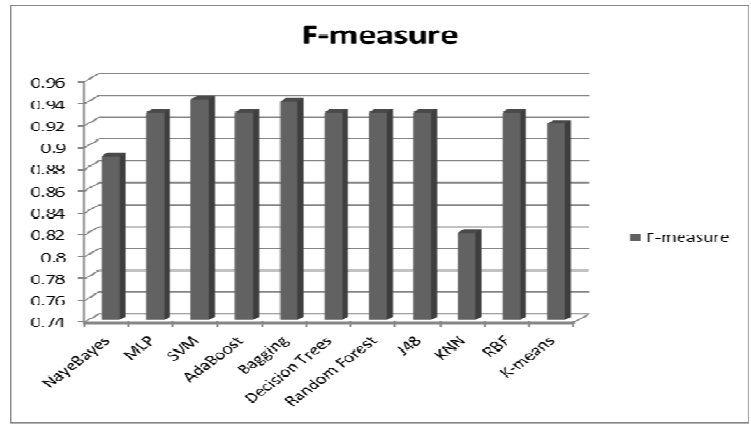

Figure 3. F-measure results for selected machine learning methods

## 4. DISCUSSION & CONCLUSION

Accuracy, F-measure and MAE results are gathered on various datasets for different algorithms as shown in Table 2, 3 & 4. The following observations were drawn from these experiment results:

*NaiveBayes* classifier for software bug classification showed a mean accuracy of various datasets 83.47. It performed really well on datasets MC1, PC2 and PC5, where the accuracy results were above 95%. The worst performance can be seen on dataset PC3, where the accuracy was less than 50%. MLP also performed well on MC1 and PC2 and got overall accuracy on various datasets 89.14 %. SVM and Bagging performed really well as compared to other machine learning methods, and got overall accuracy of around 89 %. Adaboost got accuracy of 88.59, Bagging got 89.386, Decision trees achieved accuracy around 88.47, Random Forest got 89.08, J48 got 88.33 and in the case of unsupervised learning KNN achieved 71.99, RBF achieved 88.87 and K-means achieved 87.29. MLP, SVM and Bagging performance on all the selected datasets was good as compared to other machine learning methods. The lowest accuracy was achieved by KNN method.

The best MAE achieved by SVM method which is 0.10 on various datasets and got 0.00 MAE for PC2 dataset. The worst MAE was for KNN method which was 0.27. K-means, MLP, Random Forest and J48 also got better MAE around 0.14. In the case of F-measure, higher is better. Higher F-measure was achieved by SVM and Bagging methods which were around 0.94. The worst F-measure as achieved by KNN method which was 0.82 on various datasets.

Software bugs identification at an earlier stage of software lifecycle helps in directing software quality assurance measures and also improves the management process of software. Effective bug's prediction is totally dependent on a good prediction model. This study covered the different machine learning methods that can be used for a bug's prediction. The performance of different algorithms on various software datasets was analysed. Mostly SVM, MLP and bagging techniques performed well on bug's datasets. In order to select the appropriate method for bug's prediction domain experts have to consider various factors such as the type of datasets, problem domain, uncertainty in datasets or the nature of project.

Lastly, neuro-fuzzy techniques [41-47] and software agents [48] can be used to generate test cases, increasing the efficacy of bug detection. Multiple techniques can be combined in order to get more accurate results.

## ACKNOWLEDGEMENT

The authors would like to thank Dr. Jagath Samarabandu for his constructive comments which contributed to the improvement of this article as his course work.

## AUTHORS BIOGRAPHY

**Saiqa Aleem** received her MS in Computer Science (2004) from University of Central Punjab, Pakistan and MS in Information Technology (2013) from UAEU, United Arab Emirates. Currently, she is pursuing her PhD. in software engineering from University of Western Ontario, Canada. She had many years of academic and industrial experience holding various technical positions. She is Microsoft, CompTIA, and CISCO certified professional with MCSE, MCDBA, A+ and CCNA certifications.

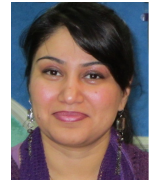

**Dr. Luiz Fernando Capretz** has vast experience in the software engineering field as practitioner, manager and educator. Before joining the University of Western Ontario (Canada), he worked at both technical and managerial levels, taught and did research on the engineering of software in Brazil, Argentina, England, Japan and the United Arab Emirates since 1981. He is currently a professor of Software Engineering and Assistant Dean (IT and e-Learning), and former Director of the Software Engineering Program at Western. His current research interests are software engineering, human aspects of software engineering, software analytics, and software engineering education. Dr. Capretz received his Ph.D. from the University of Newcastle upon Tyne (U.K.), M.Sc. from the National Institute for Space Research (INPE-Brazil), and B.Sc. from UNICAMP (Brazil). He is a senior member of IEEE, a distinguished member of the ACM, a MBTI Certified Practitioner, and a Certified Professional Engineer in Canada (P.Eng.). He can be contacted at lcapretz@uwo.ca; further information can be found at: http://www.eng.uwo.ca/people/lcapretz/

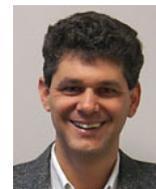

**Dr. Faheem Ahmed** received his MS (2004) and Ph.D. (2006) in Software Engineering from the Western University, London, Canada. Currently he is Associate Professor and Chair at Thompson Rivers University, Canada. Ahmed had many years of industrial experience holding various technical positions in software development organizations. During his professional career he has been actively involved in the life cycle of software development process including requirements management, system analysis and design, software development, testing, delivery and maintenance. Ahmed has authored and co-authored many peer-reviewed research articles in leading journals and conference proceedings in the area of software engineering. He is a senior member of IEEE.

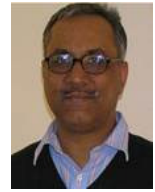